\documentclass[aps,twocolumn,groupedaddress]{revtex4}
\usepackage{amsfonts}
\usepackage{amssymb}
\usepackage[cp850]{inputenc}
\usepackage{times}
\usepackage{mathptmx}
\usepackage{graphicx}
\usepackage{epsf}

\newcommand{\be}{\begin{equation}}
\newcommand{\ee}{\end{equation}}
\newcommand{\bea}{\begin{eqnarray}}
\newcommand{\eea}{\end{eqnarray}}
\newcommand{\bi}{\begin{itemize}}
\newcommand{\ei}{\end{itemize}}

\begin{document}
\bibliographystyle{apsrev}
\title{Nanoscopic Study of the Ion Dynamics in a LiAlSiO$_4$ Glass
Ceramic by means of Electrostatic Force Spectroscopy}
\author{Bernhard Roling$^1$, Andr\'e Schirmeisen$^2$, Hartmut Bracht$^3$,
Ahmet Taskiran$^2$, Harald Fuchs$^2$, Sevi Murugavel$^1$, Frank
Natrup$^3$}
\affiliation{$^1$ Institut f\"ur Physikalische Chemie and Center
for Nanotechnology (CeNTech),
Westf\"alische Wilhelms--Universit\"at M\"unster, Corrensstr. 30, 48149 M\"unster, Germany \\
$^2$ Physikalisches Institut and CeNTech,
Westf\"alische Wilhelms-Universit\"at
M\"unster,  Wilhelm-Klemm-Str. 10, 48149 M\"unster, Germany \\
$^3$ Institut f\"ur Materialphysik and CeNTech,
Westf\"alische Wilhelms-Universit\"at M\"unster, Wilhelm-Klemm-Str. 10, 
48149 M\"unster, Germany \\}
\date{\today}
\begin{abstract}
We use time-domain electrostatic force spectroscopy (TD-EFS) for
characterising the dynamics of mobile ions in a partially crystallised
LiAlSiO$_4$ glass ceramic, and we compare the results of the TD-EFS
measurements to macroscopic electrical conductivity measurements.
While the macroscopic conductivity spectra are determined by a single
dynamic process with an activation energy of 0.72 eV, the TD-EFS
measurements provide information about two distinct relaxation processes
with different activation energies. Our results
indicate that the faster process is due to ionic movements in the glassy
phase and at the glass-crystal interfaces, while the slower process is
caused by ionic movements in the crystallites. The spatially varying electrical
relaxation strengths of the fast and of the slow process provide
information about the nano- and mesoscale structure of the glass ceramic.
\end{abstract}
\pacs{66.10.Ed, 66.30.Hs, 61.43.Fs, 61.16.Ch, 61.46.+w}

\maketitle

{\bf 1. Introduction}

Glass ceramics are important materials for a wide variety of
applications. Examples are low-expansion glass ceramics for cook
top panels and for telescope mirrors \cite{Pannhorst97,
Pannhorst04}, thin glass ceramic films for microelectronics
\cite{Shilova04} and bioactive/biocompatible glass ceramics for
bone regeneration and dental applications \cite{Zhang00,
Hoeland03}.

Low-expansion glass ceramics are based on the system Li$_2$O --
Al$_2$O$_3$ -- SiO$_2$. The low thermal expansion coefficients
arise from the negative thermal expansion coefficient of
crystallites being dispersed in a glassy matrix with a positive
thermal expansion coefficient. Furthermore, these glass ceramics
are lithium ion conductors with a conductivity depending strongly
on the degree of crystallinity. Some of the present authors have
studied the lithium ion conductivity of LiAlSiO$_4$ glass ceramics
with different degrees of crystallinity $\chi$ \cite{Roling04}.
They found that starting from a LiAlSiO$_4$ glass, partial
crystallisation leads to an enhancement of the ionic conductivity.
This effect was attributed to fast ion conduction at the
interfaces between crystallites and glassy phase. However, when
$\chi$ exceeds about 40\%, the ionic conductivity drops strongly
with increasing $\chi$, and the conductivity of the completely
crystallised ceramic is about three orders magnitude lower than
that of the glass. This suggests that the sharp drop of the
lithium ion conductivity at $\chi > 0.4$ is caused by a blocking
of the lithium ions by the poorly conducting crystallites.

Similar effects have been observed in composite electrolytes where
insulating nanoparticles are dispersed in crystalline or polymeric
ion conductors \cite{Indris00, Croce98, Croce99, Scrosati00}. At
low volume fractions of the insulating nanoparticles, the ionic
conductivity increases with increasing volume fraction, while
above a critical volume fraction, the conductivity drops with
increasing volume fraction. This strong composition dependence of
the ionic conductivity is of interest for technical applications,
since the effect can be used to prepare optimized electrolytes.
From a theoretical point of view, the mechanisms of ionic
conduction in such composite materials are not well understood. In
the case of defect crystals with dispersed insulator particles,
enhanced defect concentrations close to the interfaces between
ionic conductor and insulator particles due to space charge
effects are believed to play an important role \cite{Maier95}.
However, in structurally disordered materials, such as glasses and
polymer electrolytes, the high number density of mobile ions
implies small Debye lengths, and therefore, it seems unlikely that
space charge effects are important for ion transport. An
alternative assumption is an enhanced mobility of ions close to
the conductor--insulator interfaces.

A limiting factor hindering a better theoretical understanding is
the traditional characterization of the ion dynamics by means of
macroscopic techniques, such as conductivity spectroscopy and NMR
relaxation techniques. These techniques average generally over the
ion dynamics in different phases and at interfaces leading to a loss 
of information about the microscopic and nanoscopic
mechanisms of the ion transport. Therefore, it would be desirable
to develop experimental techniques which provide {\it spatially
resolved} dynamic information.

In a recent publication \cite{Schirmeisen04} we have shown that
time-domain electrostatic force spectroscopy (TD-EFS) using an
atomic force microscope is capable of probing ion dynamics and ion
transport in nanoscopic subvolumes of ion conducting glasses. In
the case of our experimental setup, the size of the subvolume was
about (40 nm)$^3$. Thus, the technique should be very powerful for
characterizing nano- or mesoscale structured solid electrolytes.

In this paper, we report on the application of TD-EFS to a
LiAlSiO$_4$ glass ceramic with 42\% crystallinity. At this degree
of crystallinity, the lithium ion conductivity becomes maximal
\cite{Roling04}. According to Ref. \cite{Biefeld77}, the average
size of the crystallites in this glass ceramic is about 300 nm.
Since the ionic conductivity of the crystallites is much lower
than that of the glassy phase \cite{Roling04}, this glass ceramic
is an interesting model material for obtaining a better
understanding of ion transport in nano- and mesoscale structured
solid electrolytes. The most important result of our TD-EFS
measurements is that we can distinguish two dynamic
processes, while only the faster of these two processes is
observable in the macroscopic conductivity spectra of the glass
ceramic.

\vspace{1cm}
{\bf 2. Experimental}

For the experiments we use a commercial, variable-temperature
AFM operating under ultrahigh vacuum (UHV) conditions
(Omicron VT-AFM). The sample temperature can be varied in the
range from 30 K up to 650 K. The force sensor is a single
crystalline, highly doped silicon cantilever with a resonant
frequency of 300 kHz and a spring constant of 20 N/m, featuring a sharp
conducting tip with an apex radius of 10 nm (non-contact
cantilever of type NCHR from Nanosensors). The system is operated
in the frequency modulation-mode (FM-mode) \cite{Albrecht91}, where the cantilever
is always oscillating at the resonant frequency. Conservative
tip-sample forces will induce a shift in the resonant frequency,
which is used as the feedback parameter for the
tip-sample distance control during surface scanning.

During the TD-EFS experiments, we measure the electrostatic forces
between a sample which is kept at ground potential, and a
conductive tip which is biased with a voltage of -2 V.
The electric field emanating from the tip
penetrates into the sample over a distance comparable to the tip
diameter \cite{Gomez01}. Detailed finite element simulations using
the FEMLAB software yield an approximate probed sample volume of
(40 nm)$^3$ \cite{Schirmeisen04}. If the tip bias is negative, positively
charged ions will be attracted towards the region penetrated by the electric
field. Therefore, the probed sample volume will charge up positively
compared to the surrounding region. This leads to an increase of the
electrostatic force felt by the tip and causes a negative
frequency shift. The accumulation of charged ions will continue
until an equilibrium is reached. Thus, the time-dependent change of the
frequency shift is a direct fingerprint of the dynamic conductance
behaviour of the ions.

\begin{center}
\begin{figure}[htb]
\begin{center}
\epsfxsize=8cm\leavevmode{\epsffile{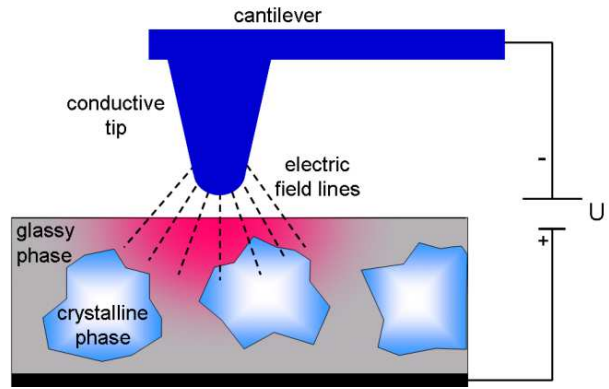}}
\caption{Schematic illustration of the experimental setup for
time-domain electrostatic force spectroscopy (TD-EFS) on a
partially crystallised glass ceramic.}
\label{fig:schematic}
\end{center}
\end{figure}
\end{center}

In Fig. \ref{fig:schematic} we sketch the basic idea behind our
TD-EFS measurements on a partially crystallised glass ceramic.
Depending on the position of the tip, the probed sample volume
contains different amounts of glassy and crystalline phases. Since
the ionic conductivity in these phases is different, we expect the
electrostatic force spectra to depend on the position of the tip.

For the preparation of a glass ceramic with 42\% crystallinity
we first prepared a glass sample as described in Ref. \cite{Roling04}.
The surface roughness of this sample was reduced to about 1-2 nm by using the
polishing techniques described in Ref. \cite{Schirmeisen04}. Subsequently,
the sample was annealed in order to generate partial crystallisation
\cite{Roling04}.
The surfaces of the resulting glass ceramic was either left untreated
or lightly polished with a water-free 1/4 micron diamond suspension
by means of a Q-tip.

The TD-EFS measurements were carried out in a temperature range from 295 K
to 607 K. The main complication during these measurements is
tem\-pera\-ture--induced drift of the tip-sample distance, especially
at elevated temperatures. To account for this effect, we first
measure the drift of the sample over a period of some minutes. The
drift rate is then calculated and during the following
spectroscopic measurements the tip position is corrected for this
drift rate.

\vspace{1cm} 
{\bf 3. Results}

\begin{center}
\begin{figure}[htb]
\begin{center}
\epsfxsize=7cm\leavevmode{\epsffile{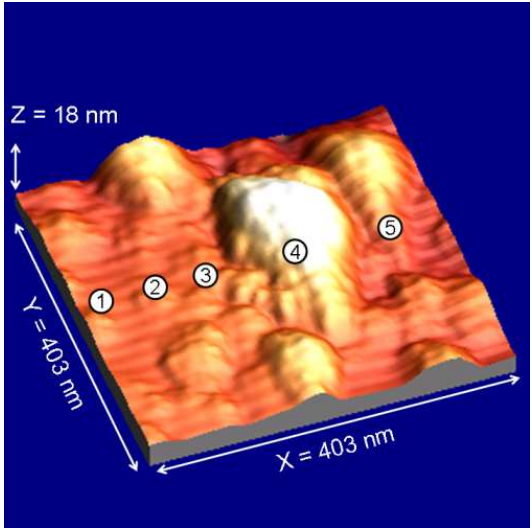}}
\caption{Surface topography of the partially crystallised glass ceramic. The
circled numbers denote tip positions during the recording of the
TD-EFS relaxation curves shown in Fig. \ref{fig:relaxcurves}.}
\label{fig:surfacescan3d}
\end{center}
\end{figure}
\end{center}

Before carrying out the TD-EFS measurements we first scanned a
small area of the sample surface. Fig.\
\ref{fig:surfacescan3d} shows a typical $400\times 400$ nm scan.
We observe protrusions on the surface with a typical diameter of $50$ to
$100$\,nm and a height varying between $5$ and $15$\,nm. It is not
clear at this point if these protrusion can be interpreted as
crystallites which grow out of the surface.

\begin{center}
\begin{figure}[htb]
\begin{center}
\epsfxsize=9cm\leavevmode{\epsffile{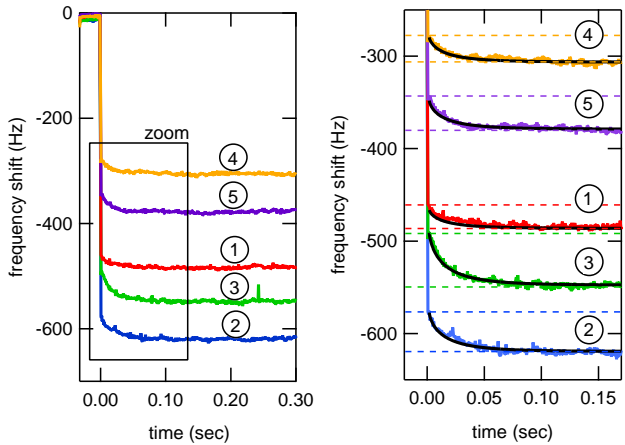}}
\caption{TD-EFS relaxation curves at a temperature of $T$ = 506 K,
obtained at the positions indicated by the circled numbers in Fig.
\ref{fig:surfacescan3d}.} 
\label{fig:relaxcurves}
\end{center}
\end{figure}
\end{center}

For each sample temperature we performed spectroscopic
measurements at different positions of the tip above the surface,
following the procedure decribed above. As an example Fig.
\ref{fig:relaxcurves} shows the
frequency shift of the oscillating cantilever as a function of
time, for $T = 506$\,K, at five different positions,
indicated by the circled numbers in Fig.\ \ref{fig:surfacescan3d}.
To start the spectroscopy measurement the distance feedback is
disabled and the tip is retracted by $1$\,nm. At time $t =
0$\,sec, a voltage of $U = -2$\,V is applied to the tip. The
relaxation curves show a sudden large negative frequency shift
$\Delta\,f_{\rm fast}$ due to fast relaxation processes.
Subsequently, slow relaxation processes are monitored until the
system has reached its saturation frequency value $\Delta\,f_{\rm
fast} + \Delta\,f_{\rm slow}$. The slow relaxation process can be fitted
with a stretched exponential function \cite{Walther98}:
\begin{equation}
\label{eq_kww} \Delta\,f(t) \;=\; (\Delta\,f_{\rm slow} -
\Delta\,f_{\rm fast}) \cdot \left[ 1\,-\,\exp(-(t/\tau)^{\beta})
\right] + \Delta\,f_{\rm fast}
\end{equation}
where $\tau$ and $\beta$ denote the temperature-dependent relaxation time
and the stretching exponent, respectively.
For the interpretation of the relaxation curves one has to
take into account that the investigated solid electrolyte consists of
crystallites embedded in a glassy matrix. From macroscopic
conductivity measurements \cite{Roling04} it is known that LiAlSiO$_4$ glass is a
moderate ion conductor with an activation energy of $E_A^{\rm glass}
= 0.72$\,eV, while a completely crystallised LiAlSiO$_4$ sample is a poor ion
conductor with a high activation energy of $E_A^{\rm crystal} = 1.07$\,eV.
At room temperature, the macroscopic electrical relaxation times
$\tau_{\rm macro} = R_{\rm macro} \cdot C_{\rm macro}$
\cite{Schirmeisen04} of the pure glass and of the completely crystallised
ceramic are about 10$^{-2}$ s and 10$^3$ s, respectively. Here,
$R_{\rm macro}$ und $C_{\rm macro}$ denote the macroscopic resistance
and capacitance, respectively.

\begin{center}
\begin{figure}[htb]
\begin{center}
\epsfxsize=9cm\leavevmode{\epsffile{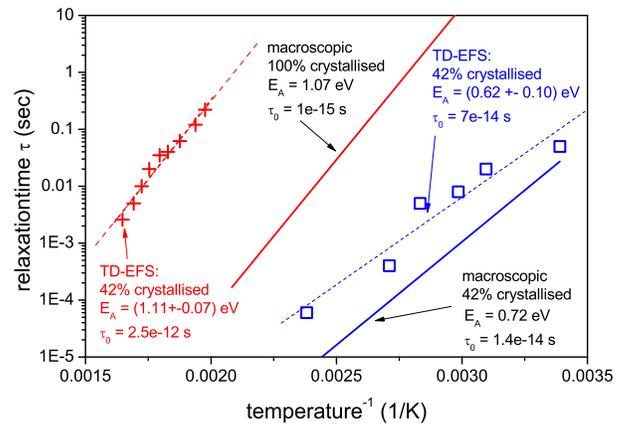}}
\caption{Arrhenius plot of TD-EFS relaxation times (symbols) and
macroscopic relaxation times (solid lines) obtained for a glass
ceramic with 42\% crystallinity and for a completely crystallised
ceramic. The dashed lines denote Arrhenius fits of the TD-EFS 
relaxation times.} 
\label{fig:arrhenius}
\end{center}
\end{figure}
\end{center}

This suggests that at room temperature, the TD-EFS relaxation curves of the
glass ceramic are determined by movements of ions in the glassy phase
and possibly at the glass-crystal interfaces \cite{Roling04}, while the ions
in the crystallites do {\it not} contribute to the relaxation. This implies
that $\Delta f_{\rm slow} = \Delta f_{\rm glass} + \Delta f_{\rm interfaces}$.
The offset $\Delta f_{\rm fast}$ is due to ultrafast vibrational and
electronic polarisations. In an Arrhenius plot in Fig.
\ref{fig:arrhenius} the relaxation times due to ionic movements
are shown as blue squares. A fit with an Arrhenius equation (dashed blue line)
yields an activation energy of (0.62 $\pm$ 0.1) eV. For comparison,
the solid blue line denotes the macroscopic electrical relaxation time with
an activation energy of 0.72 eV. Within the experimental error, the TD-EFS
and the macroscopic relaxation times are similar. However, the TD-EFS
relaxation times are characterised by a higher preexponential
factor.

At temperatures above 420 K, the relaxation process becomes so fast that
it cannot be distinguished anymore
from the ultrafast vibrational and electronic polarisation.
At temperatures above 500 K, we detect an additional slower relaxation process
with an activation energy of (1.11 $\pm$ 0.07) eV, see red crosses
(data) and dashed red line (fit) in Fig. \ref{fig:arrhenius}.
Representative relaxation curves at a temperature of T = 506 K are
shown in Fig. \ref{fig:relaxcurves}.
The solid black lines represent stretched exponential fits for
the determination of the relaxation times $\tau$. Although the
values of $\Delta f_{\rm fast}$ and $\Delta f_{\rm slow}$ vary
considerably for the different positions on the surface, the
relaxation times are consistent for the different curves.
Remarkably, the activation energy of the slow relaxation process
is almost identical to the activation energy for the macroscopic
electrical relaxation times of a completely crystallised LiAlSiO$_4$
ceramic. These macroscopic relaxation times are denoted by the solid red
line in Fig. \ref{fig:arrhenius}, corresponding to an activation energy of
1.07 eV.

\vspace{1cm}
{\bf 4. Discussion}

Our results cleary demonstrate that macroscopic averaging over the
dynamics of all ions in the glass ceramic leads to a considerable
loss of information. The macroscopic dc conductivity is determined
by the long-range ion transport along percolative diffusion
pathways. Such percolative pathways exist in the glassy phase and
along the glass--crystal interfaces \cite{Roling04}. In
comparison, the lithium ions in the crystallites move much slower,
and the crystallites do not form a percolative network
\cite{Roling04}. Therefore, the slow movements of the ions in the
crystallites do not contribute significantly to the macroscopic dc
conductivity. However, when we use TD-EFS as a local probe, we
detect two distinct relaxation processes. The activation energy of
the faster process is, within the experimental error, similar to
the activation energy of the macroscopic dc conductivity. This
suggests that this process is due to ionic movements in the glassy
phase and at the glass--crystal interfaces. The activation energy
of the slower process is similar to the activation energy for
macroscopic ionic conduction in a {\it completely crystallised}
ceramic. This indicates that the slow process we detect in our
TD-EFS measurements on the {\it partially crystallised} glass
ceramic is due to ionic movements in the crystallites. Since the
crystallites do not form a percolative network, the ionic
movements are localised and do not lead to a long-range ionic
diffusion. While such localised motions do not contribute to the
macroscopic dc conductivity, they contribute potentially to the
macroscopic ac conductivity. However, in the macroscopic ac
conductivity spectra of the glass ceramic with 42\% crystallinity,
a process with an activation energy of about 1.1 eV is not
detectable. This can be explained by the large ac conductivity
contribution of the fast ions in the glassy phase and at the
glass-crystal interfaces, which buries the ac conductivity
contribution of the slow ions in the crystallites. In contrast, by
means of TD-EFS, the slow process is detectable, when the probed
sample volume is filled to a large extent by crystallites.

An interesting question concerns the physical origin of the
discrepancies between the preexponential factors of the
macroscopic and of the TD-EFS relaxation times. In the case of the
fast process, the preexponential factor of the TD-EFS relaxation
times is slightly higher than the preexponential factor of the
macroscopic relaxation times. One possible reason lies in the
influence of the vacuum capacitance, due to the gap between tip
and sample, on the TD-EFS relaxation times, which complicates a
quantitative comparision between macroscopic and TD-EFS relaxation
times \cite{Schirmeisen04}. In the case of the slow processes, we
find that the preexponential factors of the relaxation times
differ by about three orders of magnitude. However, here we have
to take into account that the macroscopic and the TD-EFS
measurements probe different processes. The macroscopic
conductivity spectra of the completely crystallised glass ceramic
are governed by long-range ionic diffusion in a percolating
network of crystallites. The TD-EFS relaxation times of the glass
ceramic with 42\% crystallinity are determined by local movements
of ions in isolated crystallites. While the activation energies of
these distinct processes seem to be similar, it is plausible that
their preexponential factors differ considerably.

In Fig. \ref{fig:relaxcurves} we have shown that the relaxation
strengths of both the fast and the slow relaxation process depend
strongly on the position of the tip. This result can be easily
understood, since the relative amounts of glassy phase and
crystallites in the probed subvolume change when the tip moves to
a different position. A large amount of glassy phase leads to a
high relaxation strength of the faster relaxation process, while a large
amounts of crystalline phase leads to a high relaxation strength
of the slower relaxation process. Therefore, our method potentially allows us
to obtain spatially resolved spectroscopic data and to
create images of the relative contributions of the different
processes. Such images should provide comprehensive information about the
size and the distribution of the crystallites in glass ceramics.
This will be the subject of future work.

\vspace{1cm}
{\bf 5. Conclusions}

We have shown that TD-EFS measurements provide information about
ionic movements in a partially crystallised glass ceramic, which
is beyond the information obtainable by macroscopic electrical
spectroscopy. By means of TD-EFS, we have detected two distinct
relaxation processes with different activation energies. The
faster process is characterised by an activation energy similar to
the activation of the macroscopic electrical conductivity. Our
results indicate that this relaxation process is due to ionic
movements in the glassy phase and at the glass--crystal
interfaces. The slower process detected by TD-EFS does not
contribute to the macroscopic electrical conductivity of the glass
ceramic. The activation of this process is, however, similar to
the activation energy for ionic conduction in a {\it completely }
crystallised LiAlSiO$_4$ ceramic. This suggests that the slow
relaxation process in the partially crystallised glass ceramic is
caused by localised ionic movements in isolated crystallites.

The relaxation strengths of the two processes depend on the position of the
AFM tip above the surface. This reflects the spatially varying amounts
of glassy phase and crystalline phase being present in the probed subvolumes of the
sample, when the tip is moved between different positions. Thus, the TD-EFS method
can be used to obtain information on size and distribution of the crystallites
in the glassy matrix, i.e. information on the nano- and mesoscale
structure of the material.

\vspace{1cm} 
{\bf Acknowledgements}

We would like to thank the Deutsche Forschungsgemeinschaft 
and the Fonds der Chemischen Industrie for financial support
of this work.

\vspace{1cm}
{\it Based on a talk given at the 84th International Bunsen Discussion
Meeting on the 'Structure and Dynamics of Disordered Ionic
Materials' (M\"unster, Germany, October 6-8, 2004).}

\bibliographystyle{ssi}

\end{document}